\documentclass[AMA,STIX1COL]{WileyNJD-v2}
\usepackage{moreverb}
\usepackage{multirow}

\newcommand\BibTeX{{\rmfamily B\kern-.05em \textsc{i\kern-.025em b}\kern-.08em
T\kern-.1667em\lower.7ex\hbox{E}\kern-.125emX}}

\articletype{MAIN PAPER}%

\received{<day> <Month>, <year>}
\revised{<day> <Month>, <year>}
\accepted{<day> <Month>, <year>}

\begin{document}

\title{Analysis of an Incomplete Binary Outcome Dichotomized From an Underlying Continuous Variable in Clinical Trials}

\author[1]{Chenchen Ma}

\author[2]{Xin Shen}

\author[1]{Yongming Qu} 
\author[1]{Yu Du} 

\authormark{MA \textsc{et al}}

\address[1]{\orgdiv{Statistics, Data and Analytics}, \orgname{Eli Lilly and Company}, \orgaddress{\city{Indianapolis}, \state{Indiana}, \country{USA}}}

\address[2]{\orgdiv{SAS Programming}, \orgname{Brightech International}, \orgaddress{\city{Somerset}, \state{New Jersey}, \country{USA}}}

\corres{Chenchen Ma, 
Statistics, Data and Analytics, Eli Lilly and Company, Indianapolis, IN 46285, USA\\
\email{ma\_chenchen@lilly.com}}

\abstract[Abstract]{
In many clinical trials, outcomes of interest include binary-valued endpoints. It is not uncommon that a binary-valued outcome is dichotomized from a continuous outcome at a threshold of clinical interest. 
To reach the objective, common approaches include (a) fitting the generalized linear mixed model (GLMM) to the dichotomized longitudinal binary outcome and (b) imputation method (MI): imputing the missing values in the continuous outcome, dichotomizing it into a binary outcome, and then fitting the generalized linear model for the ``complete'' data. We conducted comprehensive simulation studies to compare the performance of GLMM with MI for estimating risk difference and logarithm of odds ratio between two treatment arms at the end of study. In those simulation studies, we considered a range of multivariate distribution options for the continuous outcome (including a multivariate normal distribution, a multivariate t-distribution, a multivariate log-normal distribution, and the empirical distribution from a real clinical trial data) to evaluate the robustness of the estimators to various data-generating models. Simulation results demonstrate that both methods work well under those considered distribution options, but MI is more efficient with smaller mean squared errors compared to GLMM. We further applied both the GLMM and MI to 29 phase 3 diabetes clinical trials, and found that the MI method generally led to smaller variance estimates compared to GLMM.
}
\keywords{dichotomization, estimand, generalized linear mixed model, multiple imputation, pattern mixture model, potential outcome}

\maketitle

\section{Introduction}\label{sec1}
In clinical trials, continuous outcomes are often measured at multiple time points besides baseline to form a picture of treatment effects or the progression of disease. However, binary outcomes, defined by the proportion of subjects who fall below or above a clinically relevant threshold, are frequently of clinical interest. For example, in diabetes clinical trials, clinicians not only focus on the mean of hemoglobin A1C (HbA1c) but are also interested in the proportion of subjects reaching a HbA1c target, defined by HbA1c < 7\% \cite{davies2018management} or $\le 6.5\%$. \cite{garber2013american} To analyze such data, we typically dichotomize the continuous outcome measures and then apply statistical methods based on binary outcomes.

Incomplete data are quite common in longitudinal measurements, for example, when some subjects drop out from the study early or some measurements are not obtained due to technical errors or subject's unavailability. 
Rubin \cite{rubin1976inference} classified missing mechanisms into three categories. When missingness does not depend on any values in the data set, observed or unobserved, the missing mechanism is considered missing completely at random (MCAR). A mechanism where missingness is related to observed outcomes, but not related to unobserved outcomes given the observed outcomes, is called missing at random (MAR). When missingness depends on unobserved outcomes, the mechanism is said to be missing not at random (MNAR). Both MCAR and MAR are considered as ``ignorable'' when mild regularity conditions are satisfied, since likelihood-based inferences can be made by analyzing observed outcomes without explicitly modeling the missing mechanism. In contrast, MNAR mechanism is ``non-ignorable'' under this situation.

MAR is usually a reasonable assumption in clinical trials when drawing inference from sensitivity analyses of incomplete longitudinal outcomes. \cite{molenberghs2004analyzing} The traditional approach to analyze longitudinal binary outcomes that assume MAR uses the generalized linear mixed model (GLMM), which can be easily implemented by PROC GLIMMIX in SAS. However, GLMMs utilize the dichotomized variable to model the missingness, which may not capture the full mechanism of missingness, leading to a potential information loss. In addition, GLMMs are sometimes difficult to converge or result in unstable estimates.  Alternatively, assuming MAR, one can impute the continuous outcome by the multiple imputation (MI) method first and then categorize into the binary outcome. This method may perform well intuitively because it is reasonable to believe that continuous outcomes provide more information than binary outcomes, \cite{royston2006dichotomizing} potentially resulting in more accurate imputation than indirectly imputing the dichotomized binary outcome in GLMMs. Another advantage of the MI method is that imputation strategies can be customized to create an estimator consistent with the potential outcome of interest and the corresponding estimand, \citep{ich2020e9,lipkovich2020causal} unlike the GLMM approach that provides the estimator always assume all missing potential outcomes follows the same pattern as the observed values. For example, when we assume the missing potential outcomes after certain intercurrent events will be similar to the outcomes for the \emph{retrieved dropouts}, those patients who had similar intercurrent events but without missing values, the retrieved dropout imputation may be used to impute these missing values. \cite{CHMP2010} 

Several researchers compared statistical methods to analyze the longitudinal binary outcome created from the underlying continuous outcomes assuming MAR. Lipkovich et al \cite{lipkovich2005multiple} compared the performance of multiple imputation with
generalized estimating equations and restricted pseudo-likelihood in estimating overall treatment effects and treatment differences at the last scheduled visit. Based on their simulations, with moderate to high dropout rates (40\% to 60\%),  multiple imputation led to less biased and more precise estimates of treatment differences for binary outcomes based on underlying continuous scores. In addition, Floden and Bell \cite{floden2019imputation} compared multiple imputation of the continuous and dichotomous forms of the outcome and concluded that, with a higher rate of missingness, imputing the continuous outcome was less biased and had well-controlled coverage probabilities of the 95\% confidence interval compared to imputing the binary outcome. Grobler and Lee \cite{grobler2020multiple} further compared the performance of five different imputation methods and recommended imputation using substantive-model compatible  fully conditional specification method according to their simulation studies.

All the work mentioned above related to MI are based on the assumption that the continuous outcome follows a (multivariate) normal distribution. This assumption, however, is often violated in clinical trials. In this article, we compared the performance of GLMM with MI based on the continuous outcome where the continuous outcome follows a multivariate normal distribution, a multivariate t-distribution with degree of freedom three, and a multivariate log-normal distribution, respectively,  under the MCAR or the MAR missing mechanism. Additional simulations were conducted with data resampled from a real clinical trial. Furthermore, we applied the two methods to 29 clinical trials from three diabetes compounds. Both methods were evaluated by the estimation of risk difference (RD) and logarithm of odds ratio (OR) at the last visit between treatment arms.

\section{Methods}
ICH E9 (R1) Addendum \cite{ich2020e9} provides a framework to define estimands, especially strategies to handle intercurrent events. Missing values can be either truly unobserved data or as a result of data censoring by handling intercurrent events with a hypothetical strategy. In the setting of this article, we assume the outcome of interest is the potential outcome if patients would continue the randomized treatment without intercurrent events (ie, using the controlled direct hypothetical [CDH] strategy \cite{qu2021implementation}).

Let $C_{ij}$ denote the continous outcome measurement for subject $i$ ($1 \le i \le n)$ at time point $j$ ($1 \le j \le J)$.  Without loss of generality, assume that lower values of $C_{ij}$ indicate better outcomes. Let $Y_{ij}$ denote the corresponding dicotomized binary variable such that $Y_{ij} = I(C_{ij} < \lambda)$ or $Y_{ij} = I(C_{ij} \le \lambda)$. 
Using the potential outcome introduced by Lipkovich et al., \cite{lipkovich2020causal} let $Y_{ij}(a,b)$ denote the potential outcome if a patient is initially assigned to treatment $a$ and follows a treatment $b$ after the intercurrent event. Under the CDH strategy, $a$ is always equal to $b$ and we simplify the notation as $Y_{ij}(a)$. Let $\boldsymbol{\mu}_k=(\mu_{k1}, \mu_{k2}, \ldots, \mu_{kJ})'$ denote the true mean for treatment group $k$ ($k=0$ for the control treatment and $k=1$ for the experimental treatment) under the CDH strategy such that
\[
{\mu}_{kj} = \frac{1}{n}\sum_{i = 1}^n E[Y_{ij}(k)].
\]
If $Y_{1j}(k), Y_{2j}(k), \ldots  Y_{nj}(k)$ are identically distributed, the true mean can be written as
\[
{\mu}_{kj} = E[Y_{ij}(k)].
\]
For a given subject, some of the longitudinal outcome $C_{ij}$ and $Y_{ij}$ may be missing after a certain time point. The missing data in $Y_{ij}$ can be handled through either a GLMM or by imputing the continuous outcome $C_{ij}$ using multiple imputation.

\subsection{Generalized Linear Mixed Model (GLMM)}
A GLMM for binary data takes the binomial exponential family, with canonical link being logistic. Specifically, the mean response is modeled as

$$\log\left(\frac{p_{ij}}{1-p_{ij}}\right)=\boldsymbol{X_{ij}}'\boldsymbol{\beta},$$
where $p_{ij}$ is the probability of $Y_{ij}=1$ for the $i$th subject at the timepoint $j$,  $X_{ij}$ is the covariate vectors and $\beta$ is the fixed effect. The between-subject correlation for the binary outcome can be modeled with a between-subject random effect or through the residuals in the marginal model. In this article, we will only consider the marginal model because it allows more flexible covariance structures for the within-subject correlation. The GLMM can be fitted using multiple statistical softwares, such as R, SAS, and SPSS.

\subsection{Multiple Imputation (MI) Method}
Rubin first introduced MI as a tool to handle nonresponse in large sample public use surveys. \cite{rubin1987multiple} There are three main steps in MI procedure: imputation, analysis, and pooling. In the imputation step, $m$ complete versions of the data are created by replacing the missing values with plausible data values via some chosen imputation model. Next, the analysis of interest is performed for each of the $m$ complete data sets separately. Finally, the $m$ results are pooled into a a single MI result by ``Rubin’s rules.''\cite{rubin1987multiple} The variance formula based on the within- and between-imputation variances (Rubin's rules) may over estimate the variance when the imputation model and the analysis model are uncongenial. \cite{bartlett2020bootstrap,meng1994multiple,xie2017dissecting} Therefore, we also evaluated the inference based on bootstrap methods.
 
To analyze a derived binary outcome, we can first impute its underlying continuous outcome in the imputation step, then dichotomize it into the binary outcome. Analysis and pooling steps can be further applied to the imputed data sets. By imputing the continuous outcome rather than working on the binary outcome directly, more information is expected to be utilized, which potentially improves the imputation accuracy and overall efficiency for the estimators.

\subsection{Group Means and Variance Estimation}
For GLMMs with adjustment for baseline covariates, the model-based mean for a treatment group (e.g., the least squared mean) produced by most software packages (including SAS and R) estimates the response at the mean covariate, which has little clinical meaning. Therefore, we consider the average mean probability for each treatment group and the unconditional treatment effect (RD or OR). \cite{guidancedraft} Therefore, the mean response at each treatment group should be estimated by taking the average of the predicted response for this group across all subjects from all groups.\cite{freedman2008randomization}  The marginal treatment contrasts (RD or OR for the binary outcome) can then be constructed using the group means.  
Qu and Luo \cite{qu2015estimation} further studied this problem and provided the formula for variance estimation. 

Specifically, the group means can be written as 
$$\hat{p}_k=\frac{1}{n}\sum^n_{i=1}g^{-1}(\hat{\beta}_0+X_i'\hat{\beta}_X+k\hat{\beta}_k), k=0,1,$$
where $k$ is the treatment arm indicator and $g$ is the logit link function with $g(t)=\log\{t/(1-t)\}$ and $g(y)=(1+e^{-y})^{-1}$.
It is straightforwad to obtain the  estimated group mean difference (RD) as
$$\hat{p}_1-\hat{p}_0=\frac{1}{n}\sum^n_{i=1}\left\{\frac{1}{1+e^{-(\hat{\beta}_0+X_i'\hat{\beta}_X+\hat{\beta}_k)}}-\frac{1}{1+e^{-(\hat{\beta}_0+X_i'\hat{\beta}_X)}}\right\}.$$
Using delta method, variance of risk difference can be estimated 
\begin{equation}
\label{eq1}
Var(\hat{p}_1-\hat{p}_0)=G_1^T\hat{\Sigma}_{\beta}G_1,
\end{equation}
where $\hat{\Sigma}_{\beta}$ is the estimated variance-covariance matrix for $\hat{\beta}$ (sandwich variance estimation is recommended to estimate $\hat{\Sigma}_{\beta}$) and $$G_1=\frac{1}{n}\sum^n_{i=1}\left\{\frac{e^{\hat{\beta}_0+X_i'\hat{\beta}_X+\hat{\beta}_k}}{(e^{\hat{\beta}_0+X_i'\hat{\beta}_X+\hat{\beta}_k}+1)^2}\begin{pmatrix} 1 \\ X_i \\ 1 \end{pmatrix}-\frac{e^{\hat{\beta}_0+X_i'\hat{\beta}_X}}{(e^{\hat{\beta}_0+X_i'\hat{\beta}_X}+1)^2}\begin{pmatrix} 1 \\ X_i \\ 0 \end{pmatrix}\right\}.$$
\\
Similarly, the estimated logarithm of OR [log(OR)] has the following form
$$\hat{r}_1-\hat{r}_0=\log(\frac{\hat{p}_1}{1-\hat{p}_1})-\log(\frac{\hat{p}_0}{1-\hat{p}_0})$$
with variance estimated by
\begin{equation}
\label{eq2}
Var(\hat{r}_1-\hat{r}_0)=G_2^T\hat{\Sigma}_{\beta}G_2,
\end{equation}
where $$G_2=\frac{1}{n}\sum^n_{i=1}\left\{\frac{1}{(1-\hat{\mu}_1)\hat{\mu}_1}\frac{e^{\hat{\beta}_0+X_i'\hat{\beta}_X+\hat{\beta}_k}}{(e^{\hat{\beta}_0+X_i'\hat{\beta}_X+\hat{\beta}_k}+1)^2}\begin{pmatrix} 1 \\ X_i \\ 1 \end{pmatrix}
-\frac{1}{(1-\hat{\mu}_0)\hat{\mu}_0}\frac{e^{\hat{\beta}_0+X_i'\hat{\beta}_X}}{(e^{\hat{\beta}_0+X_i'\hat{\beta}_X}+1)^2}\begin{pmatrix} 1 \\ X_i \\ 0 \end{pmatrix}\right\}.$$

\section{Simulation} \label{sec:simulation}
In this section, we compared the performance of GLMM to MI, analyzing the binary outcome derived from an underlying continuous outcome in a longitudinal setting. We considered a range of distribution options for the longitudinal continuous data. To achieve this comprehensive evaluation, we adopted two data generation strategies to obtain the complete data set with continuous values. In the first strategy, data were generated from some pre-specified distributions besides a multivariate normal distribution, including a multivariate t-distribution (heavier tails than normal) and a multivariate log-normal distribution (skewed). While in the other, data were resampled from a completed real clinical trial. We focused on the monotone missingness in our simulation. Note that a missing data pattern is said to be monotone when an outcome missing for a particular individual implies that all subsequent outcomes are missing for that individual. Missing mechanisms of MCAR and MAR were considered. The objective was to estimate the RD and log(OR) at the final visit assuming the missing potential outcome follows the same pattern as the observed data. In GLMM, the within-subject errors were modeled through the residuals using an unstructured variance-covariance
matrix. When the analysis failed to converge, an independent variance-covariance matrix would be used. For the MI method, missing values were imputed using fully conditional specification (FCS)
implemented by the MICE algorithm as described in Van Buuren and
Groothuis-Oudshoorn.\cite{van2011mice} $m=10$ completed data sets were created in the imputation step. Estimated variances were obtained by (\ref{eq1}) and (\ref{eq2}), respectively, for RD and log(OR) in each data set and pooled by ``Rubin's rules.'' Ten thousand replicates were simulated for each simulation scenario. 

All of our simulation settings were based on the IMAGINE-2 study, which was a phase 3, double-blinded, randomized diabetes clinical trial for insulin-naïve adults with type 2 diabetes. \cite{davies2016basal} The primary objective was to test for non-inferiority of peglispro to glargine for HbA1c change from baseline to week 52. As secondary objectives, the comparisons of the percentages of subjects achieving HbA1c targets of $<7\%$ and $\le 6.5\%$ were also of interest. The HbA1c was measured at baseline (week 0) and at five subsequent timepoints (week 4, week 12, week 26, week 39, and week 52). The mean HbA1c (\%) were $\boldsymbol{\mu_0}=(8.45, 7.94, 7.18, 7.07, 7.14, 7.21)$ and $\boldsymbol{\mu_1}=(8.47, 7.96, 7.07, 6.84, 6.86, 6.91)$ for the glargine (control) and the peglispro (treatment) arm, respectively. The standard deviations were very close for the two arms at each time point around $\boldsymbol{\sigma}=(1.02, 0.96, 0.79, 0.84, 0.91, 0.95)$. A total of $1518$ subjects were randomized 2:1 to the peglispro and glargine arm, respectively. Among them, $1269$ had complete HbA1c data.

\subsection{Simulation Based on Complete Continuous Data Generated From a Pre-specifed Distribution}
We considered two arms with 200 subjects in each arm and a total of six measurements on the continuous outcome of interest required for each subject.  Complete data sets for continuous outcomes, $C_{ij}$, were generated from a multivariate normal distribution, a multivariate t-distribution with degree of freedom three, and a multivariate log-normal distribution, respectively, using R packages ``mvtnorm'' and ``MethylCapSig.'' Based on empirical evidence from the IMAGINE-2 study, \cite{davies2016basal} we assumed the mean HbA1c (\%) at six measurement timepoints, including baseline, were $\boldsymbol{\mu_0}=(8.5, 7.9, 7.2, 7.1, 7.1, 7.2)$ and $\boldsymbol{\mu_1}=(8.5, 8.0, 7.1, 6.8, 6.8, 6.9)$ for the control and the treatment arm, respectively. We further assumed the standard
deviation vector over time from measurement 1 to measurement 6 to be $\boldsymbol{\sigma}=(1.02, 0.96, 0.79, 0.84, 0.91, 0.95)$ and the correlation between measurement $j$ and $k$ to be $0.8^{|j-k|}$. Given $\lambda=7 (\%)$, the binary outcomes were constructed by $Y_{ij} = I(C_{ij}< 7)$. To be consistent with the clinical data, we set the missing rates to be 3\%, 6\%, 10\%, 13\%, and 15\% across the timepoints (excluding the baseline), respectively. Under MCAR, the missingness is completely at random, ie, generated from Bernoulli distributions independent of all the other elements in subjects data vector. With MAR, the missingness depends on the longitudinal history of observed data including HbA1c and the treatment arms where the rate of dropout is higher in the control arm compared to the treatment arm and the subjects with higher HbA1c in the previous visits are more likely to dropout. The programming codes of missing data generation for MCAR and MAR are provided in Appendix \ref{ap1}.


Table \ref{t1} summarizes the empirical bias (BIAS), empirical variance (VAR), average estimated variance (EVAR), relative mean squared error (RMSE) of MI compared GLMM, and 95\% empirical coverage probability (CP) for RD and log(OR), respectively. The RMSE was calculated by the ratio of MSEs for GLMM versus MI. Note that the true values of RD were 0.125, 0.197, and 0.128 for normal, t, and log-normal distribution settings, respectively. The corresponding true values of log(OR) were 0.505, 0.801, and 0.515. Under all distribution assumptions, both GLMM and MI approaches performed well with small biases. The estimated variances were in good agreement with the empirical variances and the coverage probabilities were close to the nominal value of 95\%. Given RMSE > 1 for all scenarios, we concluded that MI provides a more efficient estimator than GLMM given those considered scenarios. By adopting ``Rubin's rules'' in the MI method, variances were overestimated resulting in higher CPs, as expected.

\begin{table}[ht]
\fontsize{7pt}{12pt}
\selectfont
\caption {Simulation Results for Data Generated From a Pre-specified Distribution} \label{t1}
\begin{center}
\begin{tabular}{ lllllllllllll } 
 \toprule
\multirow{2}{*}{DIST} & \multirow{2}{*}{MISSING} & \multirow{2}{*}{METHOD} & \multicolumn{5}{c}{RD} & \multicolumn{5}{c}{LOG(OR)}
\\
\cmidrule(l){4-8} \cmidrule(l){9-13}
&&& BIAS & VAR & EVAR  & CP & RMSE & BIAS & VAR & EVAR & CP & RMSE  \\
 \midrule
 NORMAL & MCAR & GLMM & 0.0014 & 0.0027 & 0.0026 & 0.9476 &  & 0.0088 & 0.0447 & 0.0443 & 0.9506&  \\ 
 & & MI & -0.0006 & 0.0023 & 0.0027  & 0.9621 &  1.1426 & 0.0005 & 0.0391 & 0.0444  & 0.9634& 1.1442\\
 & MAR & GLMM & -0.0001 & 0.0027 & 0.0027 & 0.9497&   & 0.0023 & 0.0452 & 0.0448 & 0.9511&  \\
 & & MI & -0.0005 & 0.0024 & 0.0027  & 0.9621 &  1.1484 & 0.0010 & 0.0394 & 0.0446  & 0.9643& 1.1470\\
 \midrule
  T(3) & MCAR & GLMM & 0.0001 & 0.0027 & 0.0029 & 0.9469 &   & 0.0052 & 0.0475 & 0.0517 & 0.9486&  \\ 
 & & MI & -0.0085 & 0.0024 & 0.0027  & 0.9580 &  1.0994 & -0.0307 & 0.0416 & 0.0467  & 0.9583& 1.1163\\
 & MAR & GLMM & -0.0017 & 0.0028 & 0.0033 & 0.9453 &  & -0.0029 & 0.0487 & 0.0587 & 0.9464&  \\
 & & MI & -0.0085 & 0.0024 & 0.0027 & 0.9562 &  1.1109 & -0.0307 & 0.0424 & 0.0468  & 0.9558& 1.1217\\
 \midrule
  LOG-NORMAL & MCAR & GLMM & 0.0014 & 0.0027 & 0.0026  & 0.9467&  & 0.0085 & 0.0444 & 0.0443 & 0.9487&  \\ 
 & & MI & -0.0007 & 0.0023 & 0.0027  & 0.9623 & 1.1336& 0.0001 & 0.0392 & 0.0443 & 0.9635& 1.1351 \\
 & MAR & GLMM & -0.0007 & 0.0027 & 0.0027 & 0.9501 &   & 0.0003 & 0.0446 & 0.0448  & 0.9513& \\
 & & MI & -0.0011 & 0.0023 & 0.0027 & 0.9644 & 1.1436  & -0.0014 & 0.0390 & 0.0446  & 0.9655&  1.1437 \\
\bottomrule
\end{tabular}
\begin{tablenotes}
   \item Abbreviations: DIST, distribution; RD, risk difference; OR, odds ratio; MCAR, missing completely at random; MAR, missing at random; BIAS, empirical bias; VAR, empirical variance; EVAR, average estimated variance; CP, 95\% empirical coverage probability; RMSE, relative mean squared error. 
  \end{tablenotes}
\end{center}
\end{table}

\subsection{Simulation Based on Resampling of Complete Continuous Data From a Diabetes Clinical Trial}
Furthermore, we conducted a simulation study with resampling from IMAGINE-2 study. \cite{davies2016basal} We considered a beneficial treatment scenario and a null effect scenario. For the beneficial treatment scenario, the mean HbA1c was lower for the treatment arm compared to the control arm. In this case, we let the mean HbA1c (\%) be $\boldsymbol{\mu_0}=(8.45, 7.94, 7.18, 7.07, 7.14, 7.21)$ and $\boldsymbol{\mu_1}=(8.47, 7.96, 7.07, 6.84, 6.86, 6.91)$ for the control and the treatment arm, respectively. Data for null effect scenario were generated by letting $\boldsymbol{\mu_0}= \boldsymbol{\mu_1}=(8.45, 7.94, 7.18, 7.07, 7.14, 7.21)$. For binary outcomes of interest, we considered two thresholds $\lambda=6.5 (\%)$ and $7 (\%)$. Corresponding binary outcomes were constructed by $Y_{ij} = I(C_{ij}\leq 6.5)$ and $Y_{ij} = I(C_{ij}< 7)$, respectively. Incomplete data sets were generated similarly as above.

The simulation process was as follows: 
(1) subset the IMAGINE-2 data with complete history of HbA1c measurements (ie, data from the 1269 subjects); 
(2) adjusted the mean response at each time point by applying a constant to each subject's outcome at that visit corresponding to $\boldsymbol{\mu_0}$ and $\boldsymbol{\mu_1}$, respectively. These served as the potential outcomes for each subject under each arm at each visit;
(3) resampled the subjects such that $n=1518$;
(4) drew treatment assignments using a Bernoulli distribution with 2:1 randomization ratio; 
(5) assigned the potential outcomes accordingly;
(6) induced dropouts according to MCAR or MAR;
(7) conducted analyses;
(8) repeated previous step (3)-(7) 10,000 times; and
(9) summarized the operating characteristics.

As in the previous section, similar conclusions could be made given the simulation results in Table \ref{t2}. Besides BIAS, VAR, EVAR, RMSE, and CP, the 95\% coverage probability based on non-parametric bootstrap with 10,000 resamples (CP-B) was reported. For the bootstrap, we calculated bootstrap percentile intervals.\cite{bartlett2020bootstrap} The true values of RD and log(OR) were 0 for null effect scenarios. For the beneficial treatment scenarios, the true values of RD were 0.128, and 0.136 for $\lambda=6.5$ and $\lambda=7$, respectively. The corresponding true values of log(RD) were 0.649, and 0.546. CP-Bs were closer to 95\% compared to CP, indicating that variance estimation was more accurate by bootstrap. These results suggested that MI may be recommended as a primary method of analysis, and corresponding variances were better estimated by bootstrap.

\begin{table}[ht]
\fontsize{7pt}{12pt}
\selectfont
\caption {Risk Difference Comparison: Data Resampled From Clinical Trial Data} \label{t2}
\begin{center}
\begin{tabular}{ llllllllllllllll } 
\toprule
\multirow{2}{*}{$\lambda$} & \multirow{2}{*}{EFFECT} & \multirow{2}{*}{MISSING} & \multirow{2}{*}{METHOD} & \multicolumn{6}{c}{RD} & \multicolumn{6}{c}{LOG(OR)}
\\
\cmidrule(l){5-10} \cmidrule(l){11-16}
&&&& BIAS & VAR & EVAR  & CP & CP-B& RMSE& BIAS & VAR & EVAR  & CP & CP-B& RMSE \\
 \midrule
 6.5& NULL & MCAR & GLMM & 0.0003 & 0.0006 & 0.0005 & 0.9453 & 0.9454&  & 0.0033 & 0.0199 & 0.0192 & 0.9460 & 0.9438& \\ 
 && & MI & 0.0003 & 0.0005 & 0.0005  & 0.9627 & 0.9463 & 1.1680& 0.0033 & 0.0172 & 0.0194  & 0.9636 & 0.9446&  1.1575\\
 && MAR & GLMM & -0.0009 & 0.0006 & 0.0006 & 0.9453 & 0.9421&  & -0.0034 & 0.0200 & 0.0195 & 0.9464 & 0.9436& \\
 && & MI & 0.0004 & 0.0005 & 0.0005& 0.9578 & 0.9440  &  1.1928 & 0.0044 & 0.0173 & 0.0191  & 0.9592 & 0.9430& 1.1598\\
 & BEN & MCAR & GLMM & 0.0018 & 0.0006 & 0.0006  & 0.9447 & 0.9453&   & 0.0135 & 0.0179 & 0.0174 & 0.9483 & 0.9446&  \\ 
 && & MI & 0.0007 & 0.0005 & 0.0006 & 0.9616 & 0.9468 &  1.1606 & 0.0095 & 0.0155 & 0.0175  & 0.9630 & 0.9453&  1.1589\\
 && MAR & GLMM & 0.0018 & 0.0006 & 0.0006 & 0.9459 & 0.9442 & & 0.0091 & 0.0179 & 0.0177 & 0.9482 & 0.9431& \\
 && & MI & 0.0005 & 0.0005 & 0.0006  & 0.9608 & 0.9487 &1.1777& 0.0081 & 0.0155 & 0.0173  & 0.9621 & 0.9444&  1.1571\\
 \midrule
 7& NULL & MCAR & GLMM & 0.0001 & 0.0008 & 0.0008 & 0.9480 & 0.9470 &   & 0.0008 & 0.0128 & 0.0125 & 0.9482 & 0.9468& \\ 
 && & MI & 0.0002 & 0.0007 & 0.0008  & 0.9607 & 0.9485 & 1.1272& 0.0009 & 0.0114 & 0.0126  & 0.9611 & 0.9482& 1.1252\\
 && MAR & GLMM & -0.0012 & 0.0008 & 0.0008 & 0.9490 & 0.9465 &  & -0.0045 & 0.0130 & 0.0127 & 0.9492 & 0.9465& \\
 && & MI & 0.0003 & 0.0007 & 0.0008 & 0.9598 & 0.9493&  1.1437 & 0.0014 & 0.0114 & 0.0125  & 0.9606 & 0.9487&  1.1386\\
 & BEN & MCAR & GLMM & 0.0016 & 0.0008 & 0.0008 & 0.9487 & 0.9485 &  & 0.0073 & 0.0126 & 0.0126 & 0.9492 & 0.9485&  \\ 
 && & MI & 0.0019 & 0.0007 & 0.0008& 0.9622 & 0.9505  &  1.1214 & 0.0084 & 0.0113 & 0.0126  & 0.9628 & 0.9502& 1.1206\\
 && MAR & GLMM & -0.0001 & 0.0008 & 0.0008 & 0.9471 & 0.9468 &  & 0.0007 & 0.0129 & 0.0128 & 0.9474 & 0.9458& \\
 && & MI & 0.0008 & 0.0007 & 0.0008  & 0.9614 &  0.9491 & 1.1381& 0.0042 & 0.0113 & 0.0127  & 0.9618 & 0.9481& 1.1360\\
 \bottomrule
\end{tabular}
\begin{tablenotes}
   \item Abbreviations: DIST, distribution; RD, risk difference; OR, odds ratio; BEN, beneficial; MCAR, missing completely at random; MAR, missing at random; BIAS, empirical bias; VAR, empirical variance; EVAR, average estimated variance; CP, 95\% empirical coverage probability; CP-B, 95\% coverage probability based on bootstrap; RMSE, relative mean squared error. 
  \end{tablenotes}
\end{center}
\end{table}

\section{Application in clinical trials} \label{sec:app}
\begin{figure}[ht]
\caption{Variance Ratio of GLMM to MI }
\centering
\includegraphics[width=1.0\textwidth]{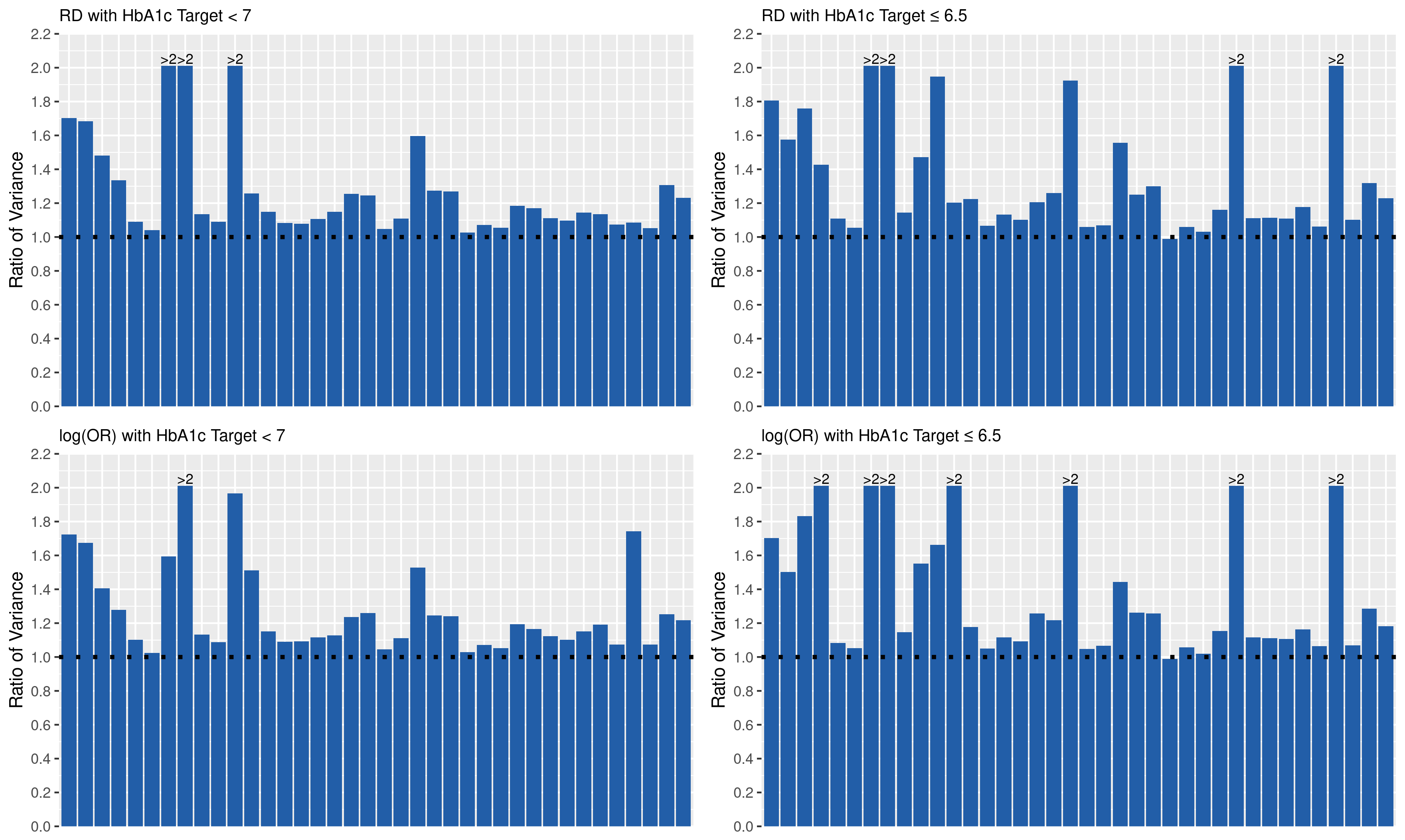}
\label{f1}
\end{figure}

\begin{figure}[ht]
\caption{Regression for Variance Ratio of GLMM to MI Versus Missing Rate}\label{f3}
\centering
\includegraphics[width=1.0\textwidth]{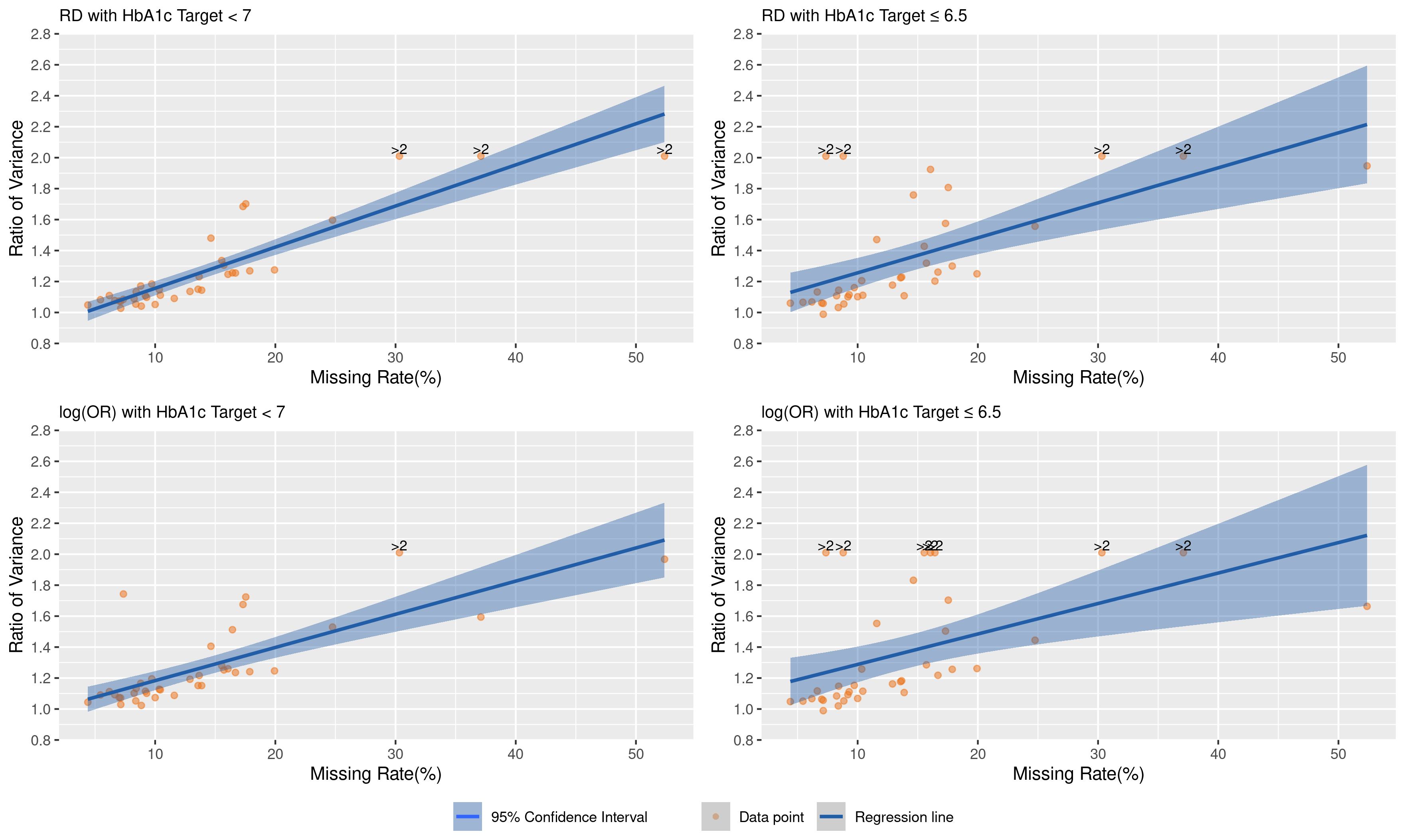}
\label{f2}
\end{figure}

In this sectioin, we applied both GLMM and MI to 29 phase 3 clinical trials for studying diabetes treatments to analyze the RD or the log(OR) in reaching HbA1c target of < 7\% or $\le$ 6.5\% at the end of the study. The durations of these study ranged from 12 to 104 weeks, and the sample sizes were between 182 and 1518. For trials with multiple treatment arms, we compared each treatment arm of interest with the reference arm, resulting in a total of 38 comparisons. Since simulation studies in Section \ref{sec:simulation} showed both methods are little biased in the estimation of the RD or the log(OR), and we were not able to estimate the bias for the real data (not knowing the truth), we focused on the relative variance (variance ratio of GLMM to MI) for these comparisons, where variances are estimated by non-parametric bootstrap. \cite{efron1994introduction} Figure \ref{f1} shows the relative variances for the RD and the log(OR) for the two treatment targets aforementioned. For the HbA1c target of < 7\%, MI had smaller variance than GLMM in all comparisons for both RD and log(OR). The mean ratios of variances between the two methods were 1.28 for both RD and log(OR), respectively. For the HbA1c target of $\le$ 6.5\%,  the MI-based approach had smaller variances than GLMM in 97.4\% (37 out of 38) of comparisons for both RD and log(OR). The mean ratios of variances between the two methods were 1.42 and 4.83 for RD and log(OR), respectively. Some ratios were high (> 2) since the GLMM was suspected to diverge with unusually high values of variance estimates for four comparisons. If we exclude those results for the four comparisons, with the HbA1c target of < 7 (\%), the mean ratio of variances were 1.20 and 1.25 for RD and log(OR), respectively; with the HbA1c target of $\le 6.5$ (\%), the mean ratio of variances were 1.27 and 1.22 for RD and log(OR), respectively. In summary, the MI approach generally led to smaller variance estimates compared to GLMM. We also performed linear regressions for the variance ratio versus the rate of missing values at the final visit (Figure \ref{f2}). Results imply that MI was more likely to obtain lower variance estimates than GLMM when the rate of missing values were higher.

\section{Summary and Discussion}
Binary endpoints derived from continuous variables are often used as the primary endpoint or important secondary endpoints in clinical trials. Missing data occur either because subjects discontinue from the study early or the outcomes collected after intercurrent events, which do not reflect the potential outcomes of interest, are censored. Therefore, it is important to understand the appropriate methods to handle the missing values for these binary endpoints. 
Literature comparing the two approaches (GLMM and MI) through simulations by generating the underlying continuous variable from a normal distribution showed MI outperforms GLMM. However, the continuous variables are often either skewed or have heavy tails and the normality assumption could be violated. Therefore, those simulation results in existing literature provided limited assurance to use the MI-based approach in practice. 

We compared GLMM with MI for analysis of the binary outcome with monotone missing data through two types of simulations. First, we compared the performance of the two methods when the underlying continuous variable was generated from normal, lognormal, and t-distributions. Secondly, we generated data by resampling data from a clinical trial in diabetes. In both simulations, MI outperformed GLMM in all scenarios. In addition, we applied both methods to 38 comparisons from 29 historical clinical trials and showed MI almost always resulted in lower variance in the estimates compared to GLMM. Finally, through the simulation studies and real data analyses, we found GLMM was more likely to fail to converge. The ratio of estimated variances for GLMM versus MI as the missing rate increases, indicating that MI would have more advantageous over GLMM as the missing rate increases.

In this article, we assumed the underlying continuous outcomes for all subjects were independently identically distributed and all missing values followed the same pattern. In real clinical trial analysis, there may be different missingness patterns according to the reasons of missingness. \cite{qu2021implementation} Compared to GLMM, MI is an easier approach to implementing such a pattern mixture model. \cite{little1993pattern} Although we did not simulate data based on a pattern mixture model, the convincing results for the MI-based approach from the uni-pattern missing model naturally suggests the missing values will be imputed well for each pattern, and a good estimation of the overall treatment effect can be obtained when there exist mixed missingness patterns.

In conclusion, to handle the missing data for binary outcomes in clinical trials, we should impute the missing values for the underlying continuous variable and then analyze the derived binary outcome. 


\bibliography{wileyNJD-AMA}

\begin{thebibliography}{10}
\providecommand \doibase [0]{http://dx.doi.org/}%

\bibitem{davies2018management}
Davies MJ, D’Alessio DA, Fradkin J, et al. Management of hyperglycemia in
  type 2 diabetes, 2018. A consensus report by the American Diabetes
  Association (ADA) and the European Association for the Study of Diabetes
  (EASD). {\it Diabetes Care} 2018\string; 41(12)\string: 2669--2701.

\bibitem{garber2013american}
Garber AJ, Abrahamson MJ, Barzilay JI, et al. American Association of Clinical
  Endocrinologists’ comprehensive diabetes management algorithm 2013
  consensus statement. {\it Endocrine Practice} 2013\string; 19\string: 1--48.

\bibitem{rubin1976inference}
Rubin DB. Inference and missing data. {\it Biometrika} 1976\string;
  63(3)\string: 581--592.

\bibitem{molenberghs2004analyzing}
Molenberghs G, Thijs H, Jansen I, et al. Analyzing incomplete longitudinal
  clinical trial data. {\it Biostatistics} 2004\string; 5(3)\string: 445--464.

\bibitem{royston2006dichotomizing}
Royston P, Altman DG, Sauerbrei W. Dichotomizing continuous predictors in
  multiple regression: a bad idea. {\it Statistics in Medicine} 2006\string;
  25(1)\string: 127--141.

\bibitem{ich2020e9}
{Committee for Medicinal Products for Human Use} . Addendum on estimands and
  sensitivity analysis in clinical trials to the guideline on statistical
  principles for clinical trials, Step 5. tech. rep., EMA/CHMP/ICH/436221/2017;
  European Medicines Agency:   2020.

\bibitem{lipkovich2020causal}
Lipkovich I, Ratitch B, Mallinckrodt CH. Causal inference and estimands in
  clinical trials. {\it Statistics in Biopharmaceutical Research} 2020\string;
  12(1)\string: 54--67.

\bibitem{CHMP2010}
{European Medicines Agency Committee for Medicinal Products for Human Use
  (CHMP)} . {\it
  \href{https://www.ema.europa.eu/en/documents/scientific-guideline/guideline-missing-data-confirmatory-clinical-trials_en.pdf}{Guideline
  on missing data in confirmatory clinical trials}}.
\newblock EMA/CPMP/EWP/1776/99 Rev.1. .
\newblock 2010.

\bibitem{lipkovich2005multiple}
Lipkovich I, Duan Y, Ahmed S. Multiple imputation compared with restricted
  pseudo-likelihood and generalized estimating equations for analysis of binary
  repeated measures in clinical studies. {\it Pharmaceutical Statistics: The
  Journal of Applied Statistics in the Pharmaceutical Industry} 2005\string;
  4(4)\string: 267--285.

\bibitem{floden2019imputation}
Floden L, Bell ML. Imputation strategies when a continuous outcome is to be
  dichotomized for responder analysis: a simulation study. {\it BMC medical
  research methodology} 2019\string; 19(1)\string: 1--11.

\bibitem{grobler2020multiple}
Grobler AC, Lee K. Multiple imputation in the presence of an incomplete binary
  variable created from an underlying continuous variable. {\it Biometrical
  Journal} 2020\string; 62(2)\string: 467--478.

\bibitem{qu2021implementation}
Qu Y, Lipkovich I. Implementation of ICH E9 (R1): A few points learned during
  the COVID-19 pandemic. {\it Therapeutic Innovation \& Regulatory Science}
  2021\string: {https://doi.org/10.1007/s43441-021-00297-6}.

\bibitem{rubin1987multiple}
Rubin DB. {\it Multiple imputation for nonresponse in surveys}.
\newblock New York, NY: John Wiley \& Sons .
\newblock 1987.

\bibitem{bartlett2020bootstrap}
Bartlett JW, Hughes RA. Bootstrap inference for multiple imputation under
  uncongeniality and misspecification. {\it Statistical Methods in Medical
  Research} 2020\string; 29(12)\string: 3533--3546.

\bibitem{meng1994multiple}
{Meng} XL. Multiple-imputation inferences with uncongenial sources of input.
  {\it Statistical Science} 1994\string; 9(4)\string: 538--558.

\bibitem{xie2017dissecting}
Xie X, Meng XL. Dissecting multiple imputation from a multi-phase inference
  perspective: what happens when God's, imputer's and analyst's models are
  uncongenial. {\it Statistica Sinica} 2017\string: 1485--1545.

\bibitem{guidancedraft}
{U.S. Food \& Drug Administration} . Draft guidance for industry, adjusting for
  covariates in randomized clinical trials for drugs and biologics with
  continuous outcomes.  2019.

\bibitem{freedman2008randomization}
Freedman DA. Randomization does not justify logistic regression. {\it
  Statistical Science} 2008\string; 23(2)\string: 237--249.

\bibitem{qu2015estimation}
Qu Y, Luo J. Estimation of group means when adjusting for covariates in
  generalized linear models. {\it Pharmaceutical statistics} 2015\string;
  14(1)\string: 56--62.

\bibitem{van2011mice}
Van~Buuren S, Groothuis-Oudshoorn K. mice: Multivariate imputation by chained
  equations in R. {\it Journal of statistical software} 2011\string;
  45(1)\string: 1--67.

\bibitem{davies2016basal}
Davies M, Russell-Jones D, Selam JL, et al. Basal insulin peglispro versus
  insulin glargine in insulin-na{\"\i}ve type 2 diabetes: IMAGINE 2 randomized
  trial. {\it Diabetes, Obesity and Metabolism} 2016\string; 18(11)\string:
  1055--1064.

\bibitem{efron1994introduction}
Efron B, Tibshirani RJ. {\it An introduction to the bootstrap}.
\newblock Boca Raton, FL: CRC press, Taylor \& Francis .
\newblock 1994.

\bibitem{little1993pattern}
Little RJA. Pattern-mixture models for multivariate incomplete data. {\it
  Journal of the American Statistical Association} 1993\string; 88(421)\string:
  125--134.

\end{thebibliography}

\appendix 
\section{Programming Codes for Missing Data Generation for MCAR and MAR}\label{ap1}
Input data include:\\
y1--y6: complete data vector at timepoints 1--6; \\
data: complete dataset;\\
trt: treatment indicator.

\subsection{MCAR}
 R1 <- rbinom(n, 1, 0.97)\\
 R2 <- rep(0, n)\\
R2[R1 == 1] <- rbinom(sum(R1), 1, (1 - 0.031))\\
R3 <- rep(0, n)\\
R3[R2 == 1] <- rbinom(sum(R2), 1, (1 - 0.042))\\
R4 <- rep(0, n)\\
R4[R3 == 1] <- rbinom(sum(R3), 1, (1 - 0.033))\\
R5 <- rep(0, n)\\
R5[R4 == 1] <- rbinom(sum(R4), 1, (1 - 0.023))\\
data <- data \%>\% mutate(y2 = ifelse(R1 == 1, y2, NA), y3 = ifelse(R2 == 1, y3, NA), y4 = ifelse(R3 == 1, y4, NA), y5 = ifelse(R4 == 1, y5, NA), y6 = ifelse(R5 == 1, y6, NA))
        
\subsection{MAR}
p1 <- with(data, expit(logit(0.99) - 0.12 * y1 * trt - 0.14 * y1 * (1 - trt)))\\
R1 <- rbinom(n, 1, p1)\\
R2 <- rep(0, n)\\
p2 <- with(data[R1 == 1, ], expit(logit(1 - 0.031) - 0.5 * residuals(lm(y2 $\sim$ y1)) * trt - 0.7 * residuals(lm(y2 $\sim$ y1)) * (1 - trt)))\\
R2[R1 == 1] <- rbinom(sum(R1), 1, p2)\\
R3 <- rep(0, n)\\
p3 <- with(data[R2 == 1, ], expit(logit(1 - 0.042) - 0.51 * residuals(lm(y3 $\sim$ y1)) * trt - 0.72 * residuals(lm(y3 $\sim$ y1)) * (1 - trt)))\\
R3[R2 == 1] <- rbinom(sum(R2), 1, p3)\\
R4 <- rep(0, n)\\
p4 = with(data[R3 == 1, ], expit(logit(1 - 0.033) - 0.52 * residuals(lm(y4 $\sim$ y1)) * trt - 0.74 * residuals(lm(y4 $\sim$ y1)) * (1 - trt)))\\
R4[R3 == 1] <- rbinom(sum(R3), 1, p4)\\
R5 <- rep(0, n)\\
p5 <- with(data[R4 == 1, ], expit(logit(1 - 0.023) - 0.53 * residuals(lm(y5 $\sim$ y1)) * trt - 0.76 * residuals(lm(y5 $\sim$ y1)) * (1 - trt)))\\
R5[R4 == 1] <- rbinom(sum(R4), 1, p5)\\
data <- data \%>\% mutate(y2 = ifelse(R1 == 1, y2, NA), y3 = ifelse(R2 == 1, y3, NA), y4 = ifelse(R3 == 1, y4, NA), y5 = ifelse(R4 == 1, y5, NA), y6 = ifelse(R5 == 1, y6, NA))

\end{document}